\documentclass[conference]{IEEEtran}
\IEEEoverridecommandlockouts
\usepackage{cite}
\usepackage{amsmath,amssymb,amsfonts}
\usepackage{algorithmic}
\usepackage{graphicx}
\usepackage{graphics} 
\usepackage{multirow}
\usepackage{cite}
\usepackage{verbatim}
\usepackage{url}
\usepackage{ctable}
\usepackage[normalem]{ulem}
\usepackage{amsmath,amssymb,amsfonts}
\usepackage{algorithmic}
\usepackage{graphicx}
\usepackage{textcomp}
\usepackage{xcolor}
\usepackage{tabularx}
\usepackage{makecell}
\usepackage{mathtools}
\usepackage{commath}
\usepackage{array,multirow}
\usepackage{textcomp}
\usepackage{xcolor}

\usepackage{cite}
\usepackage{amsmath,amssymb,amsfonts}
\usepackage{algorithmic}
\usepackage{graphicx}
\usepackage{textcomp}
\usepackage{xcolor}

\def\BibTeX{{\rm B\kern-.05em{\sc i\kern-.025em b}\kern-.08em
    T\kern-.1667em\lower.7ex\hbox{E}\kern-.125emX}}
        
\usepackage{fancyhdr}
\begin{document}

\title{Classification of Cognitive Load and Expertise for Adaptive Simulation using Deep Multitask Learning\\
\thanks{This work was supported by the Innovation for Defence Excellence and Security (IDEaS) program, Canada.}
}

\author{\IEEEauthorblockN{Pritam Sarkar$^1$, Kyle Ross$^1$, Aaron J. Ruberto$^2$, Dirk Rodenburg$^3$, Paul Hungler$^4$, Ali Etemad$^1$ }
\IEEEauthorblockA{
$^1$\textit{Department of Electrical and Computer Engineering}\\ 
$^2$\textit{Departments of Emergency Medicine and Critical Care Medicine} \\
$^3$\textit{Faculty of Engineering and Applied Science} \\
$^4$\textit{Department of Chemical Engineering} \\
\textit{Queen's University, Kingston, Canada} \\
\{pritam.sarkar, 12kjr1, 13ajr3, djr08, paul.hungler, ali.etemad\}@queensu.ca}}


\maketitle
\thispagestyle{fancy}

\begin{abstract}
Simulations are a pedagogical means of enabling a risk-free way for healthcare practitioners to learn, maintain, or enhance their knowledge and skills. Such simulations should provide an optimum amount of cognitive load to the learner and be tailored to their levels of expertise. However, most current simulations are a one-type-fits-all tool used to train different learners regardless of their existing skills, expertise, and ability to handle cognitive load. To address this problem, we propose an end-to-end framework for a trauma simulation that actively classifies a participant's level of cognitive load and expertise for the development of a dynamically adaptive simulation. To facilitate this solution, trauma simulations were developed for the collection of electrocardiogram (ECG) signals of both novice and expert practitioners. A multitask deep neural network was developed to utilize this data and classify high and low cognitive load, as well as expert and novice participants. A leave-one-subject-out (LOSO) validation was used to evaluate the effectiveness of our model, achieving an accuracy of $89.4\%$ and $96.6\%$ for classification of cognitive load and expertise, respectively.
\end{abstract}


\section{Introduction} \label{introduction}
Simulation has been shown to be a highly effective pedagogical strategy and has been widely adopted within the healthcare industry \cite{motola_2013, good_2003}. Not only can simulation be used to aid novice healthcare professionals in becoming experts it can also help in the continual development of fundamental skills \cite{mehler_2012, Aggarwali_2010}. For trauma medicine, a fundamental skill required for responders is the management of their cognitive load. 

Exceeding cognitive capacity has been shown to significantly degrade medical performance \cite{Rodenburg_2018, Fraser_2015, Merriënboer_2010}. Cognitive load has been found to be inversely correlated with level of expertise \cite{Kalyuga_2003}. To become expert trauma responders, novice trainees need a way to learn to deal with intense cognitive load without any risks to the patient. 

Cognitive load is comprised of mental load and mental effort. Mental load is imposed based on the amount of information given whereas mental effort is the mental capacity that must be allocated to the information \cite{Paas_1992}. Experts are able to ignore extraneous details and information decreasing both their mental load and mental capacity. This, in turn, reduces the cognitive load of experts compared to novices \cite{Kalyuga_2003}. 

It has previously been shown that learning is impaired when a task exceeds the learner's cognitive load or when the learner's cognitive load is exceedingly low \cite{Brunken_2010, Teigen_1994}. If simulations do not match the expertise and cognitive load of the participant, the simulations cannot achieve the desired learning outcomes \cite{Sweller_1994}. Consequently, to increase efficacy, simulation environments should match the level of expertise and cognitive load of each learner. 

Biometric measures have been shown to indicate cognitive load \cite{Paas_1994}. Variations in cognitive load can be captured with metrics quantified from ECG measurements \cite{Sahadat_2013}. Therefore, since the link between cognitive load and expertise \cite{Paas_2003} has previously been established, we hypothesize that ECG can be utilized for accurate classification of expertise. To the best of our knowledge, no work has been done on attempting to classify the level of expertise of healthcare practitioners utilizing ECG for training simulations. 

\begin{figure*}[t]
    \centering
    \includegraphics[width=0.85\linewidth]{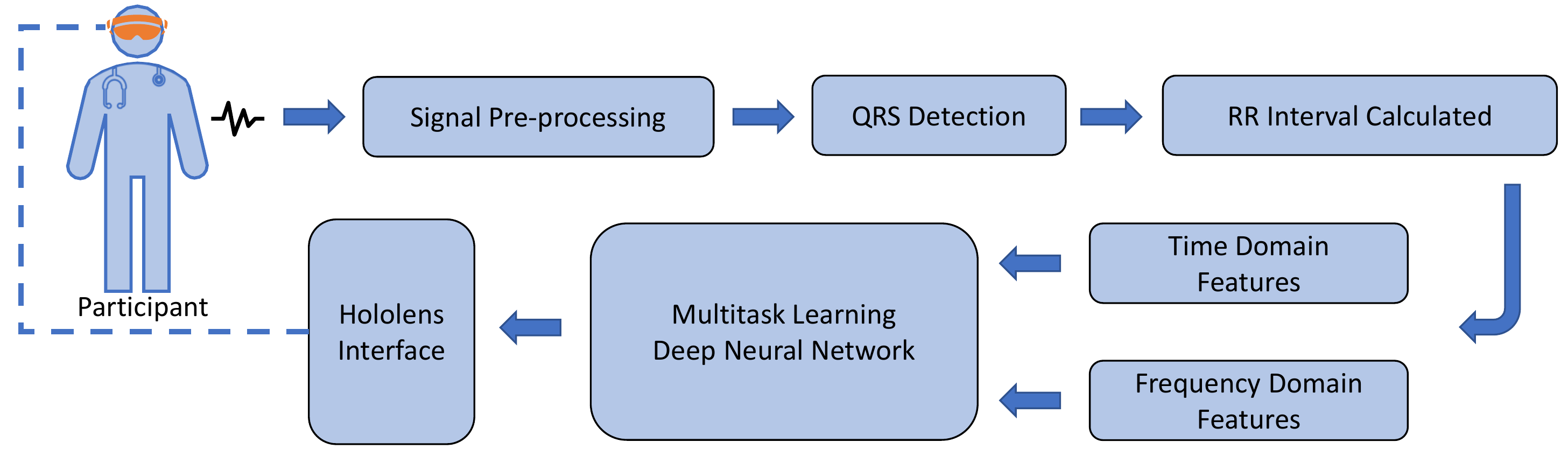}
    \caption{The proposed architecture is presented. Simulations were designed for trauma responders tending to a critically injured patient (mannequin). A Microsoft HoloLens was used to place augmented reality objects in the simulation room and to capture first-person video during simulations. The ECG data were captured from participants. Pre-processing was performed, followed by QRS detection, RR interval calculation, and feature extraction. A deep multitask neural network was then developed to classify the extracted features into expert and novice, as well as high vs. low cognitive load. The outcome can then be utilized (in future work) to modulate the simulation based on the attributes of the trainee.}
    \label{fig:arch}
\end{figure*}

In this paper, we create a novel framework for an adaptive simulation for training trauma responders, capable of dynamically adapting to the cognitive load and the level of expertise of the learner (see Figure \ref{fig:arch}). In our proposed framework, augmented reality (AR) enhancements will be utilized to modulate certain aspects of the simulation based on the cognitive and mental state of the user. To support this framework, we designed and performed two sets of trauma simulations for the acquisition of cognitive load and expertise data from trauma responders. The ECG data were collected using a wearable device, along with quantitative self-reported values of cognitive load. Expertise levels were designated based on past training and education history. Next, collected ECG data was pre-processed and time and frequency domain features were extracted. We then developed a deep multitask neural network capable of accurately classifying high vs. low cognitive load as well as expert vs. novice participants simultaneously. A leave-one-subject-out (LOSO) validation was used to evaluate the effectiveness of our model on new participants. Our proposed architecture showed great performance when compared to other works in the field, achieving an accuracy of $89.4\%$ and $96.6\%$ for classification of cognitive load and expertise, respectively. 

The rest of this paper is organized as follows. In Section \ref{background}, we discuss the related work on classification of cognitive load as well as other forms of affect using bio-signals. Next, in Section \ref{method}, we describe the experiment and simulation setup, data collection protocol, feature extraction techniques, and our proposed deep neural network used for the classification of cognitive load and expertise. In Section \ref{results}, we present the results of our method and compare the performance to other works in the field. In this Section, we also analyze the results of the two classified attributes with respect to one another, followed by a discussion on the limitations and potential future work. Lastly, Section \ref{conclusion} concludes the paper.

\section{Background} \label{background}
Affective computing using physiological signals has received considerable attention in recent years \cite{picard2001toward, healey2005detecting, liu2009dynamic, zhai2006stress}. While this field has mostly focused on applications such as driver monitoring \cite{healey2005detecting, barua2017classifying}, mental health \cite{liu2009dynamic, jerritta2011physiological}, and multimedia quality of experience \cite{Gabana_2017, navarathna2017estimating}, adaptive and dynamic training and simulation is another area where affective computing has great potential. For example, in \cite{conati2002probabilistic}, a probabilistic model for monitoring the user's level of engagement during a computer-based educational game was proposed. By analyzing ECG, electromyogram (EMG), and galvanic skin response (GSR) through a dynamic decision network, it was illustrated that bio-signals can be used to generate interactions tailored to both the user's learning and emotional state.

The types of affective states often studied in the field range from emotions \cite{picard2001toward, liu2009dynamic, jerritta2011physiological} such as stress and happiness/sadness to cognitive load \cite{healey2005detecting, barua2017classifying} and level of expertise \cite{kirby2014wireless, oviatt2018dynamic, zhou2014combining}. While analysis of emotions conform to the classical arousal vs. valence model, cognitive load and expertise analysis are not necessarily placed on the circumplex of emotions \cite{plutchik1997circumplex}. Instead, these states are often correlated with how people experience and handle arousal and valence, and therefore, the same type of approaches used for emotional states can be used for such studies.

Analysis of emotions using bio-signals for stress detection during driving was performed in \cite{healey2005detecting}. Physiological signals were collected from $24$ participants who went through $50$-minute to $1.5$-hour driving tasks. ECG, EMG, skin conductivity, and GSR along with gaze data were collected during the study. Features were then extracted from segments of $5$ minute intervals, and $3$ different stress levels were detected using linear discriminant analysis, achieving an accuracy of $97.3\%$. Additionally, through analysis of continuous features calculated at $1$-second intervals, it was shown that skin conductivity and heart rate metrics are highly correlated with each other. In another work in this area, high and low emotional valence and arousal were investigated based on electroencephalogram (EEG), EMG, and GSR \cite{girardi2017emotion}. The study was performed on $19$ subjects where $8$ videos were shown to participants in $4$ trials. Each trial consisted of a $30$-second video clip as a baseline, followed by $2$ minutes of music videos corresponding to low/high, arousal/valence states. Finally, classification was performed using a support vector machine (SVM) classifier with a polynomial kernel. In the case of arousal detection, a combination of EEG and GSR data showed the best performance, achieving an F1-score of $0.638$, whereas, in the case of valence, the best performance was achieved using all the sensor data (EEG, GSR, EMG), reporting an F1-score of $0.585$. To provide a gaming experience tailored to individual players, an affect-based dynamic difficulty adjustment (DDA) mechanism was developed in \cite{liu2009dynamic}. In the proposed setup, a player's physiological signals, including ECG, EMG, skin conductivity, and body temperature, were analyzed to infer anxiety levels. The difficulty of the game was adjusted in real time according to the player's affective state. Furthermore, a comparative study of regression tree (RT), \textit{k}-nearest neighbour (\textit{k}NN), and Bayesian network (BN) was performed, and showed that the SVM outperformed the other approaches, achieving an accuracy of $88.9\%$.

Toward automated analysis and classification of cognitive load, a number of studies have been previously conducted. In \cite{oschlies2017preliminary}, cognitive load was examined in the context of human computer interactions. This dataset was collected from $40$ participants who went through a series of mental tasks, with $6$ varying difficulty levels, to test their cognitive reaction with respect to difficulty. ECG, EMG, GSR, and body temperature were collected in order to perform the classification task. Performance comparison of $3$ different machine learning classifiers, \textit{k}NN, naive Bayes (NB), and random forest (RF), was presented using LOSO and $10$-fold validation techniques. It was found that RF outperformed the other two classifiers with a reported accuracy of $57.84\%$. In another study, classification of drivers' cognitive load was performed \cite{barua2017classifying} using an EEG dataset, where data were collected from $33$ participants in a high-fidelity moving-base driving simulator. The tasks were performed in three different simulated driving scenarios: hidden exit, crossing, and side wind. A case-based reasoning (CBR) classifier was designed based on extracted time domain and frequency domain features. The highest accuracy achieved was $70\%$ while performing hidden exit scenario. 

Lastly, toward classification of expertise, the use of bio-signals have been quite rare, to the best of our knowledge. Instead, analysis of expertise based on handwriting has been explored \cite{oviatt2018dynamic, zhou2014combining}. In \cite{zhou2014combining}, high-school students went through a task of solving mathematical problems, while a digital pen was used for recording data. Features were then extracted through analysis of pen strokes on the solution sheet. Then, three different machine learning classifiers, namely SVM, RF, and NB, were applied, reporting an accuracy of $66.7\%$, $70.8\%$, $66.7\%$, respectively. In \cite{kirby2014wireless}, a method was proposed to assess the performance of trainee surgeons during fundamental laparoscopic surgery tasks such as peg transfer, precision cutting, intracorp knot, and ligating loop. The data were captured using a $3$-axis accelerometer and a $3$-axis gyroscope. An unsupervised machine learning algorithm, \textit{k}-means clustering, was then performed on the motion data to analyze the correlation between hand movements and level of expertise.

\section{Method} \label{method}
In this section, the method including simulation design, data collection protocol, pre-processing, feature extraction, and machine learning techniques used for classification of expertise and cognitive load are presented.

\subsection{Experiments and Data} 
\begin{figure}
    \centering
    \includegraphics[width = 1\linewidth]{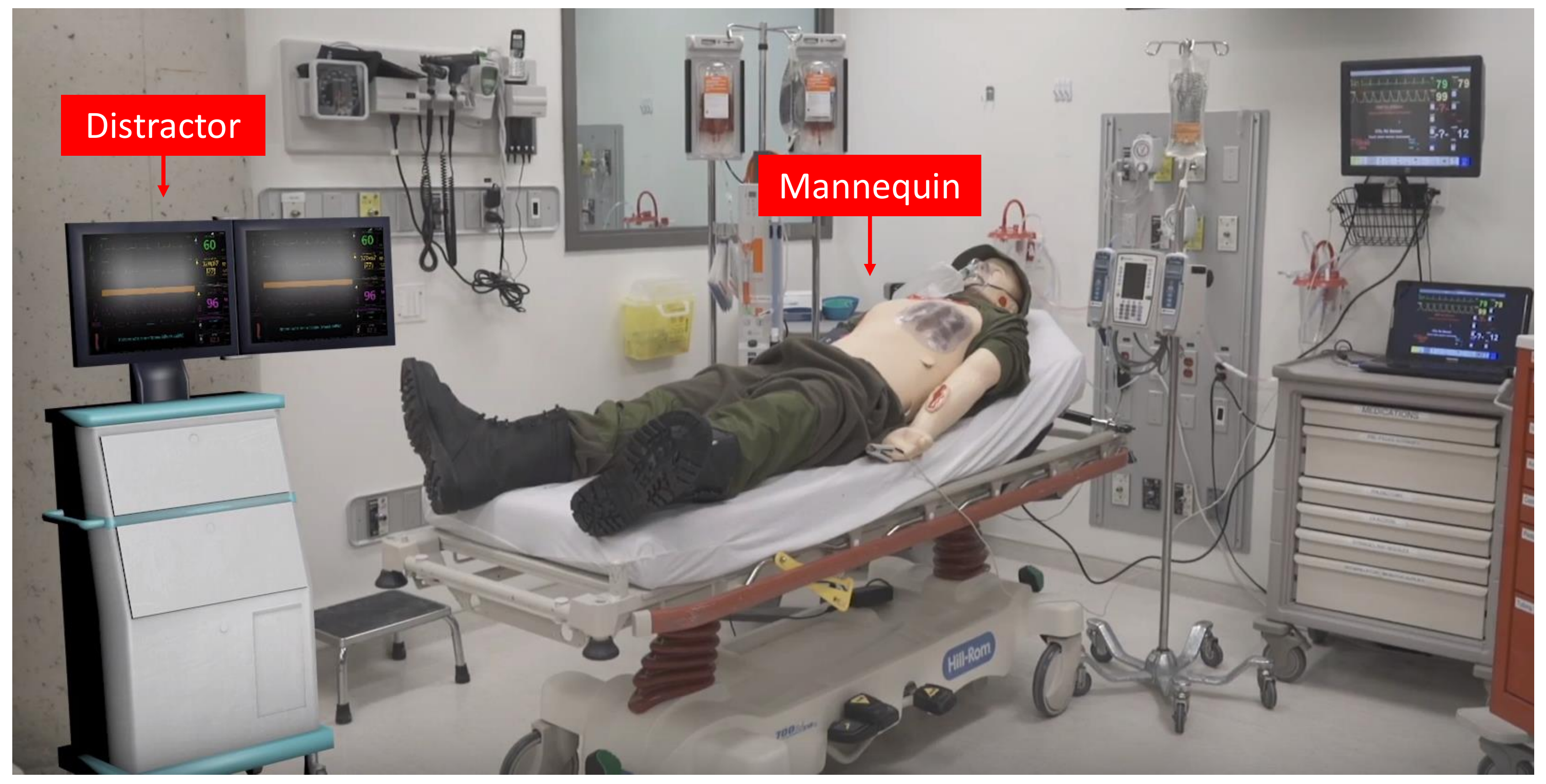}
    \caption{The simulation room and setup is presented, including the mannequin. The AR distractor shown was introduced as part of the enhanced version of our trauma simulation.}
    \label{fig:setup}
\end{figure}

Two separate trauma simulations were developed for the collection of ECG data.  
In one simulation the patient had suffered a gunshot wound to the abdomen. This was referred to as the \textit{Penetrating Trauma Simulation}. In the other simulation, the patient had been involved in an automobile roll over, referred to as the \textit{Blunt Force Trauma Simulation}. Two versions of each simulation were created, basic and enhanced. The enhanced version utilized a Microsoft HoloLens, worn by the participants, to place augmented reality objects in the simulation room to add visual complexity and distractors to the scenario \cite{HOLO}, whereas the basic version of the simulations did not use any visual enhancement. Figure \ref{fig:setup} shows the simulation room setup with patient monitor (distractor) displaying superfluous information to participants as part of the enhanced version of the simulation. First-person video was recorded from the HoloLens' front facing camera for use in debrief sessions. The goal of the simulation was for participants to monitor and tend to the patient throughout the duration of the simulation.

A Shimmer3 ECG wearable sensor \cite{shimmer} was used in this study for collecting $3$-channel ECG data. Five electrodes were attached to the chest, while the electronic module was attached to the waist for the purpose of mobility. The ECG data were captured at a sampling rate of $500$ Hz and streamed to Matlab for the purpose of live visualization and recording.

Ethics approval was secured from the Research Ethics Board of Queen's University, Canada. There were a total of $9$ participants, where $5$ were physicians specially trained in emergency medicine (considered \textit{expert} participants). This group had an average age of $34.8$ and standard deviation of $2.31$. $4$ of the participants were Queen's University medical students at the end of their $4$\textsuperscript{th} year of medical studies (considered \textit{novice} participants) with an average age of $28.5$ and standard deviation of $3.77$. Nonetheless, the students had some exposure to trauma medicine as they had been rotated through multiple medical specialties (i.e. internal medicine, surgery, emergency medicine, etc.). The reason for selecting students with some background in trauma medicine as novices was so that they could complete the aforementioned simulation.

The novice and expert participants were randomly divided into two groups. The first group performed the \textit{basic} blunt force trauma simulation followed by the \textit{enhanced} version of the penetrating trauma simulation. The second group performed the \textit{basic} penetrating trauma simulation then the \textit{enhanced} blunt force trauma simulation. Each simulation was designed for $10$ minutes in duration. Two minutes of ECG data was captured prior to the simulation to be used as \textit{baseline data}, for normalization purposes. At the end of each simulation, participants went through a debrief session where high and low cognitive load was first explained to them. Next, their first-person video was shown, and a measure of cognitive load was recorded for critical events throughout the simulation on a $1$ to $9$ scale. 


\subsection{ECG Pre-processing}
\subsubsection{Electrocardiography} 
\begin{figure}
    \centering
    \includegraphics[width = 1\columnwidth]{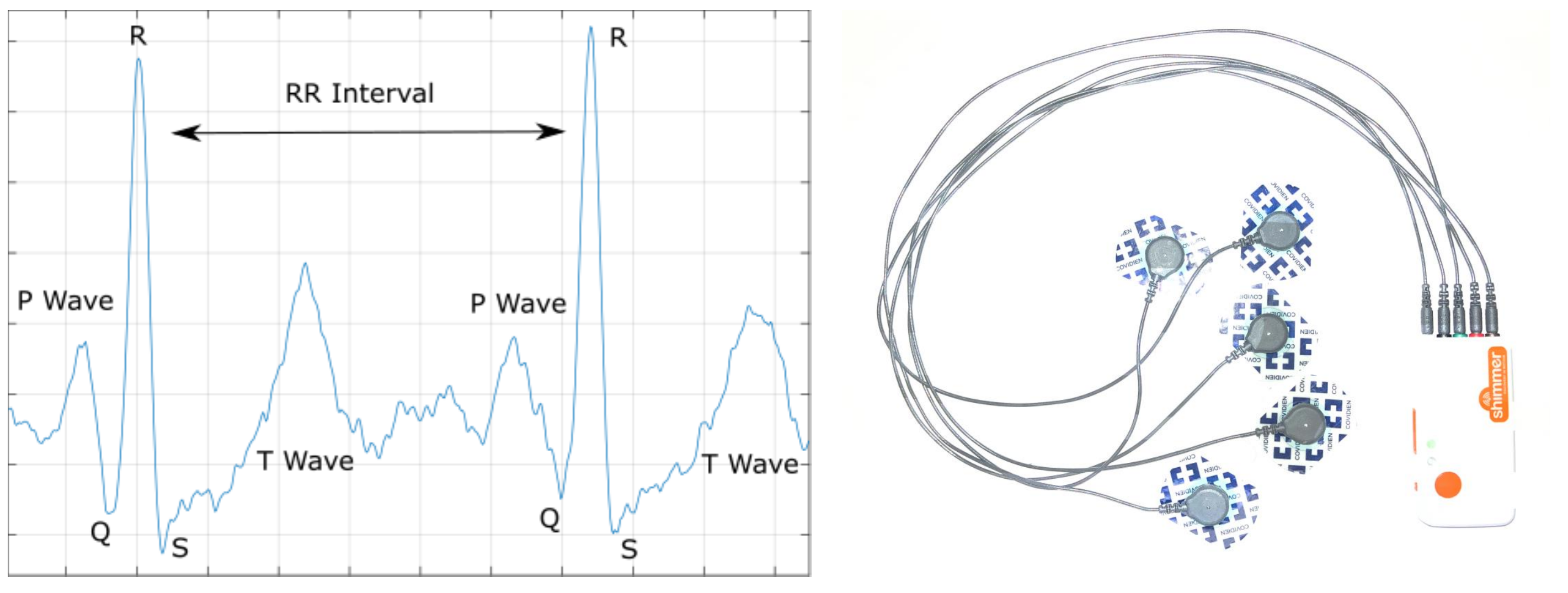}
    \caption {A sample ECG segment (left) and the wearable ECG monitoring device used in this study (right) are presented.}
    \label{fig:ecg}
\end{figure}

ECG signals were recorded using a $5$-electrode ($4$ leads and $1$ ground) wearable ECG System.  ECG signals capture the variations within cardiac electrical potentials over time, showing quasi-periodic behaviour comprised of a sequence of heart beats. Each beat is composed of three waves, a P wave, the QRS complex, and a T wave, as shown in Figure \ref{fig:ecg}. The P wave represents atrial depolarization, the QRS complex represents ventricular depolarization, and the T wave represents ventricular re-polarization \cite{ashley2004cardiology}. QRS detection acts as the starting point for ECG feature extraction. Through detecting consecutive R-peaks (the positive peaks of QRS complexes), RR intervals can be extracted. RR intervals themselves can then be used for analysis of heart rate variability (HRV), which provides significant information regarding cardiovascular behaviour, which can be influenced by factors such as health and affective/mental states of the subjects.

\subsubsection{Pan-Tompkins algorithm} The raw ECG signals were processed using the Pan-Tompkins (PT) algorithm \cite{Pan-Tompkins, hussein2018automated}. A Butterworth band-pass filter, with a passband frequency of $5-15$ Hz, was used to remove muscle noise, powerline noise, baseline wander, and T-wave interference. After noise removal, the output was differentiated to determine the slope information of the QRS complex. The absolute value of the signal was taken and a moving average filter was used to obtain wave form features in addition to the R-peaks. Filtered signals were segmented in $10$-second windows with $5$ seconds of overlap. A thresholding technique was then applied to detect R-peaks, which were further used to extract time-domain and frequency-domain features.

\subsection{Features}
Features were extracted from the identified RR intervals. All features were extracted from both the baseline data and simulation data. The features from the simulation data were normalized with respect to the baseline data. The following sections describe the two categories of features extracted successive to normalization.

\subsubsection{Time domain features} First, statistical features were extracted from the RR intervals. $11$ time domain features \cite{healey2005detecting,camm1996heart} were extracted as follows: minimum (RR\textsubscript{min}), maximum (RR\textsubscript{max}), difference between RR\textsubscript{max} and RR\textsubscript{min} (RR\textsubscript{diff}), mean (RR\textsubscript{mean}), standard deviation (RR\textsubscript{SD}), and coefficient of variation (RR\textsubscript{CV}). In addition, the RMS value of the squared differences of RR intervals (RMSSD) and standard deviation of successive differences (SDSD) were calculated. To understand short-term heart rate variability changes, the number of RR intervals greater than $50$ ms (NN$50$) was calculated, along with the percentage of NN$50$ (PNN$50$). The average heart rate (HR) of each time window was also calculated.

\subsubsection{Frequency domain feature} In addition to time domain features, $9$ frequency domain features were extracted \cite{camm1996heart, healey2005detecting}. A Lomb periodogram \cite{lomb1976least} technique was used for power spectral density (PSD) analysis. PSD estimation provides the basic information of how the power of the signal is distributed as a function of frequency. The features were extracted using the band power of different ranges of frequencies, such as ultra low frequency (ULF) band $(<0.003$ Hz$)$, very low frequency (VLF) band $(0.003$ Hz$ - 0.04$ Hz$)$, low frequency (LF) band $(0.04$ Hz$ - 0.15$ Hz$)$, high frequency (HF) band $(0.15$ Hz$- 0.4$ Hz$)$, and total power $(0$ Hz$- 0.4$ Hz$)$ of the power spectrum were calculated. Furthermore, $4$ relative frequency domain features were calculated, namely, normalized LF, normalized HF, ratio of low to high frequency power (LF/HF), and the sum of low and mid-frequency (MF) range normalized by the high-frequency range ((LF+MF)/HF), where MF is defined as $(0.08$ Hz$-0.15$ Hz$)$. The features calculated in the frequency domain are highly indicative of the sympathetic and parasympathetic systems. For example, LF is often associated with the sympathetic nervous system, while HF is associated with the parasympathetic system, and LF/HF corresponds to the overall balance between sympathetic and parasympathetic systems \cite{camm1996heart}.

\subsection{Classification}
A deep learning approach \cite{lecun2015deep} was used for classification of expertise and cognitive load. Our proposed model was based on a feed-forward neural network. Generally, when attempting to classify different attributes using neural networks, two approaches can be employed: (\textit{i}) single-task learning, where each attribute is learned individually using a separate model. This is often used as the default method for classification; and (\textit{ii}) \textit{multitask learning}, where the multiple attributes to be classified are learned simultaneously using the same model \cite{zhang2014facial, li2016deepsaliency, zhang2016joint}, sharing several hidden layers prior to calculating separate loss functions and performing the classification task for each attribute. Through this approach and by forcing the classifier to learn the internal structures of the related tasks, multitask learning performs better in cases where the learnable tasks possess some interconnections. As a result, given the probable relationship between expertise and the way in which experts and novices handle and manage cognitive load, we utilized a multitask learning approach in this work.

\begin{figure}[t]
    \centering
    \includegraphics[width = 1\linewidth]{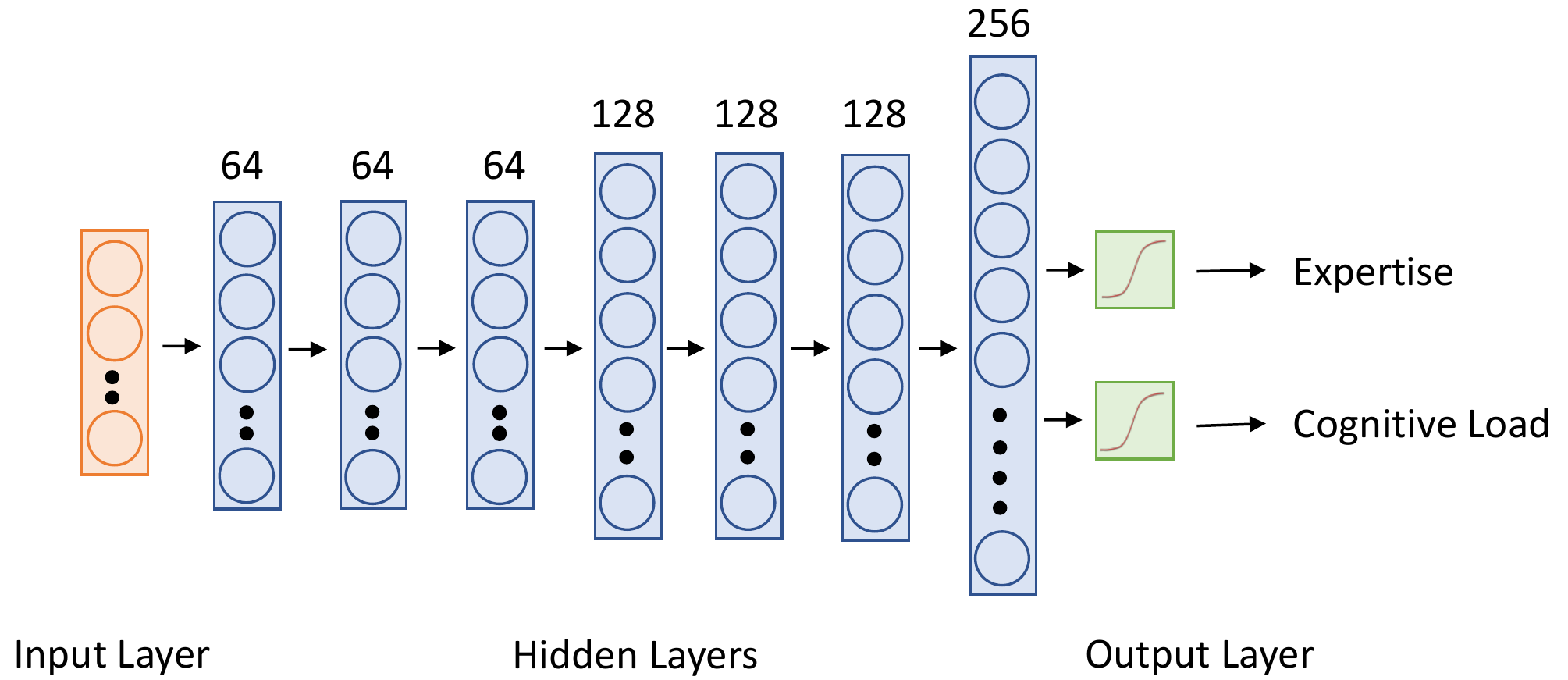}
    \caption{Our proposed deep multitask neural network architecture is presented.}
    \label{fig:network}
\end{figure}

Our proposed deep neural network (as shown in Figure \ref{fig:network}) was created with $7$ hidden layers, the first $3$ layers consisting of $64$ neurons, followed by another $3$ layers containing $128$ neurons each. The final hidden layer consisted of $256$ neurons. After each dense hidden layer, a leaky ReLu activation function was used along with dropout to overcome overfitting during training. The last dense layer was connected to two parallel sigmoid output layers in order to classify expertise and cognitive load simultaneously. The network parameters (hidden layers, hidden neurons, dropout, and activation functions) were all set empirically. Slightly different parameter values, for example $5$ or $6$ hidden layers or slightly smaller number of hidden neurons, also resulted in good a performance. However, the selected parameters resulted in a marginally better outcome. To determine the estimator function, cross-entropy loss was used:
\begin{equation}
    \begin{aligned}
        L = -\frac{1}{N}\sum_{i=1}^{N}{[y_i \log P(y_i) + (1 - y_i) \log (1 - P(y_i))]},
    \end{aligned}
    \label{equ:equ1}
\end{equation}
where $y$ is the label and $P(y)$ is the predicted probability of novice/expert or high/low cognitive load for all $N$ points. For training the model, a loss function that combines the two individual losses and minimizes both at the same time was used. Our approach considers equal weights of both individual losses:
\begin{equation}
\begin{aligned}
L_{total} = L_{expertise}+L_{cognitive\ load}
\end{aligned}
\end{equation}

The Adam Optimizer \cite{kingma2014adam} (a stochastic optimization method) was used with a learning rate of $0.001$, to train the neural network. During training process, neurons sometimes become mutually dependent in fully-connected layers, resulting in the network overfitting to the training data. To prevent the occurrence of this phenomena, a dropout rate of $50\%$ was used after each fully-connected layer so that during training, certain sets of units are not considered during a particular set of forward and backward propagation steps. Additionally, a $L2$ regularizer was introduced to overcome overfitting during training. 

\section{Results and Discussions} \label{results}
\textbf{\textit{Implementation:}} Successive to extraction of the features, our proposed classification architecture was implemented using Tensorflow \cite{tensorflow2015-whitepaper} on a NVIDIA 1070 Ti GPU. Our multitask deep neural network described above was trained with a batch size of $64$, for $400$ epochs with a learning rate of $0.001$. To evaluate the performance of our model on new subjects, we used the LOSO validation scheme. 

\textbf{\textit{Performance:}} The results of our proposed architecture are presented in Table \ref{table:results}. Our model was able to achieve an average accuracy of $96.48\%$ and $88.74\%$ for classification of level of expertise and cognitive load, respectively. To further measure the performance of our method, precision, recall, negative predictive value (NPV), and $F1$-score were also calculated. These parameters showed the high classification performance of our proposed deep multitask neural network and its ability to perform well on new participant data.

\begin{figure}
    \centering
    \includegraphics[width = 1\linewidth]{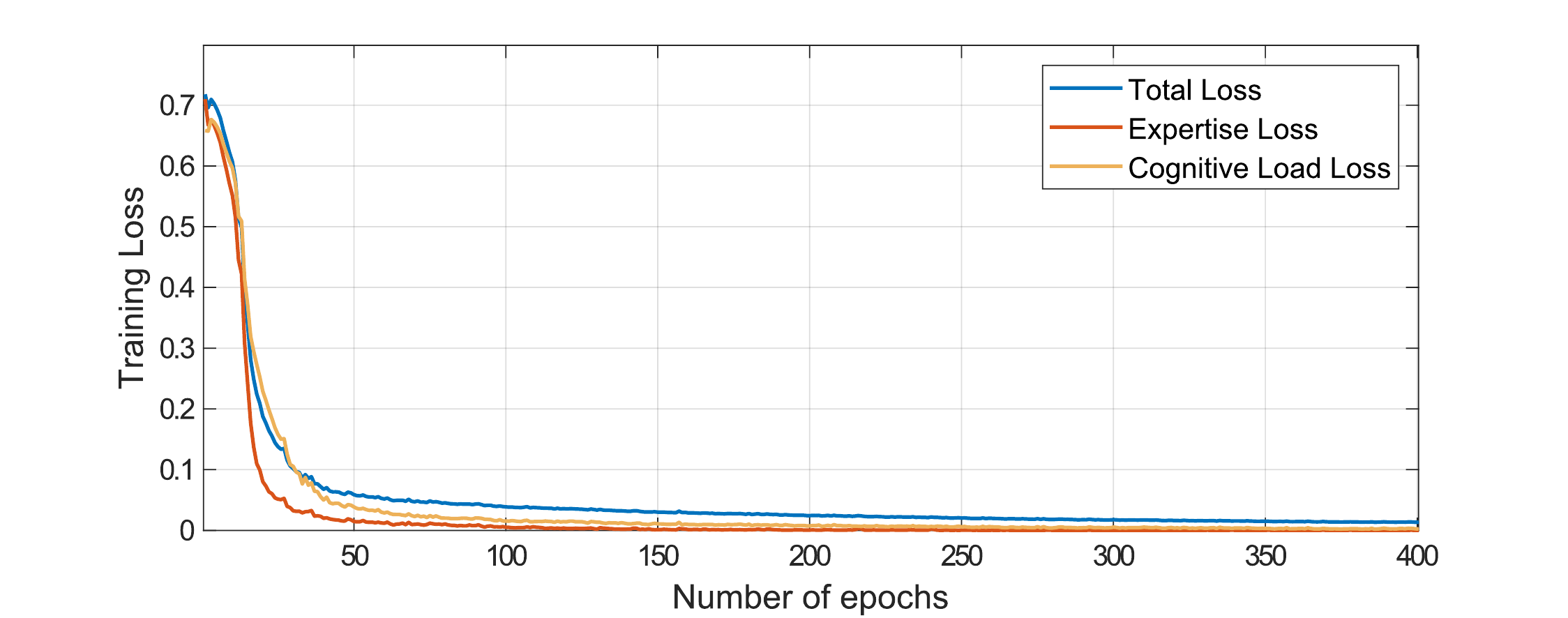}
    \caption{The training losses for expertise and cognitive load, as well as the total loss vs. the number of epochs are presented.}
    \label{fig:loss}
\end{figure}

In Figure \ref{fig:loss} we demonstrate the training losses of expertise and cognitive load against the number of epochs by averaging all the experiments of the LOSO validation scheme. Additionally, the total loss calculated as the sum of losses of expertise and cognitive load is plotted. The figure shows that the cognitive load classification loss is slightly higher than expertise classification loss, which can also be confirmed by the higher accuracy of expertise classification versus cognitive load. However, both the losses closely follow each other, showing stable training through the multitask approach. Training was performed for $400$ epochs, as it remained steady after a few hundred epochs. To further investigate and experiment with the training process of our proposed model, we removed the dropout layers and the $L_2$ regularizer. Overfitting was observed in the absence of dropout and regularization, resulting in low test accuracy.

\begin{table}
\centering
\caption{The results of our method using the LOSO scheme.} \label{table:results}
\vspace*{-3mm}
\begin{center}
\footnotesize
\centering
\begin{tabularx}{\columnwidth}{cYYYYY}
\Xhline{2\arrayrulewidth} & Accuracy & Precision & Recall & NPV & F1-score \\ \hline\hline
Expertise & 96.6 & 97.7 & 95.4 & 95.6 & 0.965 \\
Cog. Load & 89.4 & 95.7 & 81.8 & 85.0 & 0.882 \\
\Xhline{2\arrayrulewidth}
\end{tabularx}
\end{center}
\end{table}

\textbf{\textit{Comparison:}} In Table \ref{tab:result}, we compare our results to prior studies on classification of cognitive load while we could not find any other works attempting to automatically classify expertise. Though there are some literature on classification of cognitive load based on different biological signals like EEG, ECG, and GSR, in our study, we only used ECG, given our goal of achieving a simple and low-cost wearable solution. Accordingly, we decided not to compare our results to studies that had utilized EEG due to its high price and difficulty in integration with AR systems. Nonetheless, other bio-signals such as EMG, GSR, and body temperature are good candidates for future integration into our developed system and provide a good baseline for evaluating the accuracy of our model. 

The table illustrates that our proposed model performs robustly when compared to several other studies in the field. In \cite{oschlies2017preliminary}, cognitive load classification was performed based on ECG, EMG, GSR, and body temperature captured during a question answering task. Classification was performed using three machine learning classifier namely \textit{k}NN with $k=5$, an RF with $11$ trees, and an NB with $11$ kernels with a minimum bandwidth of $0.1$. A LOSO validation accuracy of $50.4\%$, $56.3\%$, and $57.8\%$ were achieved for \textit{k}NN, NB, and RF respectively. It is evident that our model outperforms this study by a considerable margin. However, it should be noted that three classes of cognitive load (low, medium, and high) were used in \cite{oschlies2017preliminary}. Next, to further evaluate our model, in the absence of prominent literature on automatic classification of cognitive load and expertise, we have compared our work to well-cited literature in another field of affective computing, anxiety/stress. The reason that we have selected this attribute for comparison is the fact that there has been a clear relationship established between stress and cognitive load \cite{setz2010discriminating}. It can be seen in Table \ref{tab:result} that our work performs well compared to the mentioned studies.

\begin{table}[t]

    \centering
    \caption{Comparison of our proposed deep multitask neural network (DMNN) with previous approaches.}
    \begin{tabular}{|c c c c|c c|}
    \hline            
         \textbf{Ref.} & \textbf{Task}  & \textbf{Attribute} & \textbf{Signals} & \textbf{Method} & \textbf{Acc.} \\ 
        
         \hline\hline
          \multirow{3}{*}{\cite{oschlies2017preliminary}} & \multirow{3}{*} {\makecell{Mental\\Task}} & \multirow{3}{*} {\makecell{Cog. Load}} & \multirow{3}{*}{\makecell{ECG, EMG,\\GSR, Temp}} & $k$NN & $50.4\%$ \\
          \cline{5-6} &&&& NB & $56.3\%$ \\
          \cline{5-6} &&&& RF & $57.8\%$ \\
        \specialrule{.1em}{.0em}{.0em}   
         \multirow{4}{*}{\cite{liu2009dynamic}} & \multirow{4}{*} {\makecell{Computer\\Game}} & \multirow{4}{*}{Anxiety} & \multirow{4}{*}{\makecell{ECG, GSR,\\Temp}} & $k$NN & $80.4\%$ \\
          \cline{5-6} &&&& BN & {$80.6\%$} \\
          \cline{5-6} &&&& RT & {$80.4\%$} \\
          \cline{5-6} &&&& SVM & {$88.9\%$} \\
         \specialrule{.1em}{.0em}{.0em}
          {\cite{healey2005detecting}} & \makecell{Driving\\task} & Stress & \makecell{ECG, \\EMG, GSR} & {LDA} & $97.3\%$ \\
        \specialrule{.1em}{.0em}{.0em}   
         \multirow{2}{*}{\cite{setz2010discriminating}} & \multirow{2}{*}{\makecell{Arithmetic\\Task}}& \multirow{2}{*}{Stress} & \multirow{2}{*}{\makecell{GSR}} & SVM & $81.3\%$ \\
          \cline{5-6} &&&& LDA & {$82.8\%$} \\
        \specialrule{.1em}{.0em}{.0em}
          \multirow{2}{*}{\textbf{\makecell{Ours}}} & \multirow{2}{*}{\makecell{Training\\Simulation}} & Expertise & \multirow{2}{*}{ECG} & \multirow{2}{*}{DMNN} & {$96.6\%$} \\ & & Cog. Load & &  &  {$89.4\%$} \\
          \hline             
    \end{tabular}
    
    \label{tab:result}
\end{table}

\begin{figure}
    \centering
    \includegraphics[width = 0.8\linewidth]{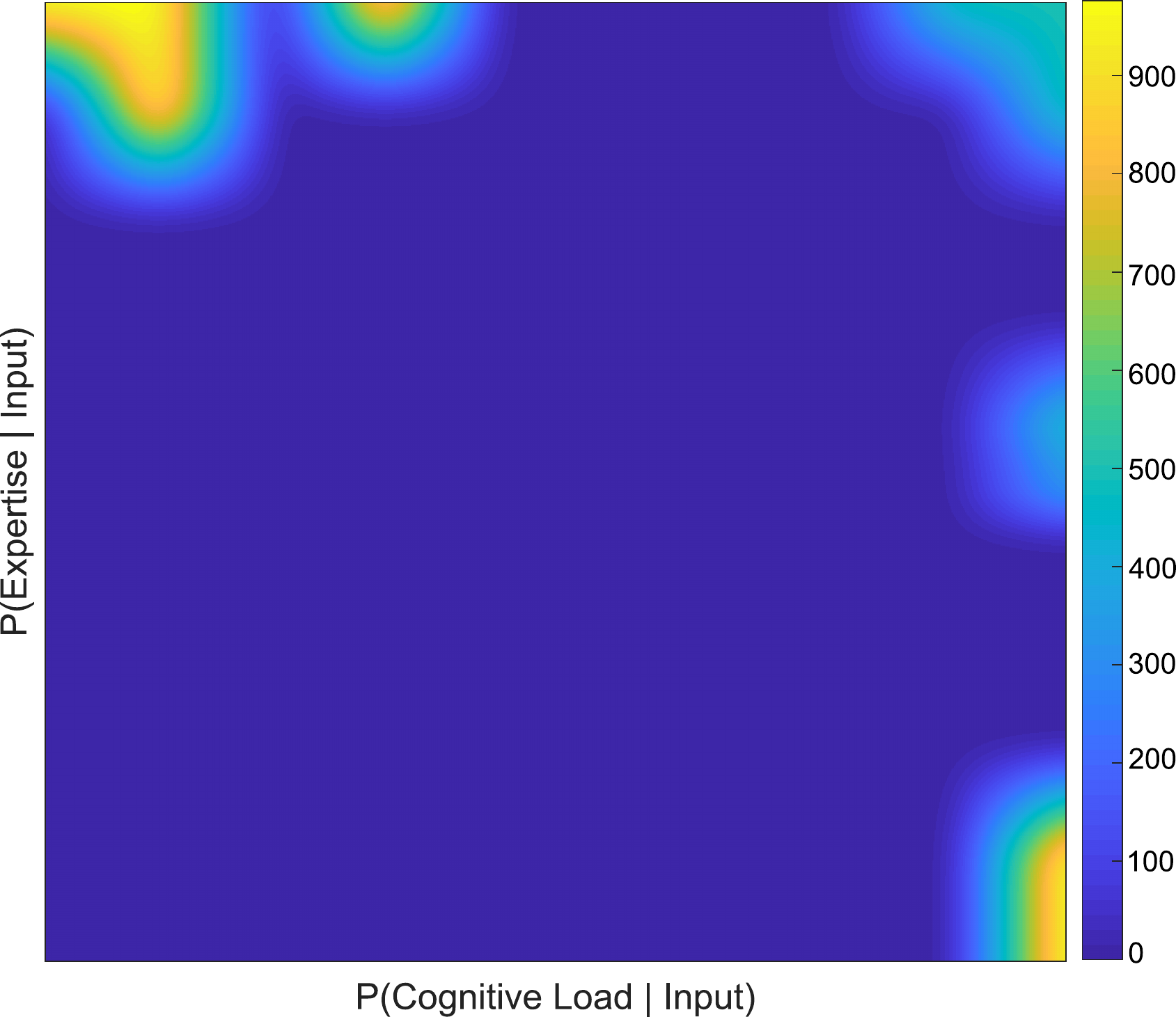}
    \caption{The output probabilities for cognitive load vs. expertise using our model for each input feature-vector is presented. Brighter areas indicate larger number of produced outputs. A relatively inverse relationship is observed. The colorbar indicates the concentration of the data points.}
    \label{fig:analysis}
\end{figure}

\textbf{\textit{Cognitive Load vs. Expertise:}} Further investigations showed an interesting relationship between participants' level of expertise and their cognitive load. In Figure \ref{fig:analysis}, the output of our model for each of the two attributes given every input feature-vector is presented. The probabilities of these attribute are then used as \textit{x} and \textit{y} coordinates to derive Figure \ref{fig:analysis}. As shown in the figure, the outputs form two clusters around $[0,1]$ and $[1,0]$. This distribution points to the fact that our model has learned a relatively inverse relationship between cognitive load and expertise, inferring that throughout our simulation, experts generally are expected to experience less cognitive load (or handle cognitive load better than novices), while novice participants experience higher amounts of cognitive load. The figure does show some output data along the expertise axis (e.g. around $= 0.5$) for high predicted cognitive load, indicating that in some cases, higher values of cognitive load can also be experienced by expert subjects.

\textbf{\textit{Limitations and Future Work:}} One of the limitations of our study is the number of subjects ($9$) making up the dataset. While the amount of data recorded for these participants was sufficient to successfully train and test our proposed multitask neural network, we believe further data will aid in a more accurate and robust classifier. Furthermore, electrodermal activity is known to provide important information for measuring cognitive load and other forms of affect \cite{healey2005detecting, setz2010discriminating}. In our future work and during the next round of data collection, GSR will be recorded along with ECG, which will allow for additional affect-related features to be extracted and learned by classifiers. Lastly, as part of our future work, adaptive components will be added to the simulation through the AR headset. The algorithms developed in this study will be used to estimate the expertise and cognitive load of participants, which will, in turn, modify certain components of the training simulation. The impact of such dynamic and adaptive changes will then be monitored and studied for gaining insights into the impact of fine-tuned simulation for trauma training purposes.

\section{Conclusions} \label{conclusion}
We proposed and developed a pipeline for facilitating an increase in educational efficacy through the development of adaptive simulation, dynamically tailored to the learner's level of expertise and cognitive load. We tackled this problem through classification of cognitive load and expertise. ECG signals were recorded from expert and novice trauma responders during two trauma simulations. Feature extraction was performed in both time and frequency domains for use in a deep learning classifier. A deep multitask neural network was developed and a LOSO validation scheme was used to evaluate the accuracy of our model. Average accuracies of $96.6\%$ and $89.4\%$, and $F1$ scores of $0.965$ and $0.882$ were achieved for classification of level of expertise and cognitive load, respectively, showing great performance when compared to other works in the field. The results show the feasibility of accurate classification of expertise and cognitive load using wearable devices. Our proposed framework can enable an increase in educational efficacy through the development of adaptive simulations, dynamically tailored to the learner's level of expertise and cognitive load.

\bibliography{reference.bib}
\bibliographystyle{ieeetr}

\end{document}